# Sensor Control for Multi-Object Tracking Using Labeled Multi-Bernoullie Filter

Amirali K. Gostar, Reza Hoseinnezhad and Alireza Bab-Hadiashar
School of Aerospace, Mechanical and Manufacturing Engineering
RMIT University
Victoria 3083, Australia

*Abstract*—The recently developed labeled multi-Bernoulli (LMB) filter uses better approximations in its update step, compared to the unlabeled multi-Bernoulli filters, and more importantly, it provides us with not only the estimates for the number of targets and their states, but also with labels for existing tracks. This paper presents a novel sensor-control method to be used for optimal multi-target tracking within the LMB filter. The proposed method uses a task-driven cost function in which both the state estimation errors and cardinality estimation errors are taken into consideration. Simulation results demonstrate that the proposed method can successfully guide a mobile sensor in a challenging multi-target tracking scenario.

## I. INTRODUCTION

In the context of multi-target tracking, the aim of sensor-control is to direct sensor(s) (by applying a set of admissible control commands) toward an unknown number of targets to maximize observability. In general, each control command positions sensor(s) to a new state(s) which results in generating different sets of measurements. Each set of the generated measurements contains information which differ from other sets. The generated information can be analyzed via a decision making process (e.g. optimizing an objective function) and as a result, the right control command could be determined in order to maximize the utility of the measurements.

The complexity of this procedure is caused by uncertainty embedded in both state and measurement spaces. In control theory, such problems are addressed by the stochastic control theory in which the number of targets may vary randomly when the time evolves. Also the observation is affected by noise, false alarm or miss detection. A natural choice to model sensor control problem is the *Partially Observed Markov Decision Processes* (POMDPs) framework in which an observer (e.g. mobile sensor) cannot reliably identify the underlying actual state (e.g. target states).

Recently, the finite set statistics (*FISST*) [1] has received substantial attention to address the underlying state estimation process in POMDP framework [2]–[12]. FISST is based on considering the multi-target entity in both state and measurement spaces as a random finite set (RFS). Several solutions for multi-target tracking problems have been proposed and implemented in FISST framework, such as the *PHD* [1], *CPHD* [13], *MeMBer* [1], *CB-MeMBer* [14], Labeled Multi-Bernoulli (*LMB*) [15] and its general version δ-Generalized Labeled Multi-Bernoulli (*δ-GLMB*) [16] filters.

In FISST-based sensor control framework, a criterion is defined to evaluate the *quality* of the updated multi-target density after a control command is applied to the sensor. In this approach, the control command is chosen to provide the *best* updated density based on the defined criterion. Example of this approach is employing the *Csiszár* in Mahler's solution for sensor-control problem [17]. He utilized the Csiszár as the reward function within a FISST filtering scheme. However, he later introduced a new reward function and forged it as the "*Posterior Expected Number of Targets*" (PENT) [4], [5].

The commonly used criterion to evaluate the quality of the updated distribution is its *divergence* from the predicted distribution [7]. In two consecutive papers, Ristic et al. [7], [8] used Rényi divergence as a *reward function* to quantify the information gained via updating the predicted density using sensor-control technique. In those works, Mahler's FISST [1] was used as the framework for multi-target Bayesian filtering. In [7], the implementation of Rényi reward maximization was investigated for the general form of multi-target filters with random finite set (RFS) assumptions for the multi-target state. Since this approach is computationally intractable even for a small number of targets [8], in the second paper [8] the PHD-based filter was used to propagate the multi-object posterior, which facilitates approximation of the Rényi divergence function via i.i.d. assumption.

Recently, a number of solutions have been developed for sensor control within a multi-Bernoulli filter [9]–[11], [18]. In these works, new task-driven objective functions are defined and optimized, as a result of which, sensor control is aimed to *directly* minimize cardinality and state estimate errors. This approach is in stark contrast to sensor control with information driven objective functions (such as Rényi divergence) where the enhancement in quality of measurements is expected to be resulted from gaining the most informative posterior density. Gostar et al. [11] defined a new objective function for the sensor-control problem in the multi-Bernoulli filter framework. This objective function is based on the statistical mean of cardinality variance in conjunction with state estimate errors. In a similar work, Hoang [18] used the "MAP" cardinality variance of the multi-Bernoulli filter.

In this paper we propose an alternative approach for the solution of multi-target sensor control problem by exploiting a new family of the RFS and its related filter. The Labeled Multi-Bernoulli RFS (which is the special case of δ-GLMB RFS) is a new family of RFS which conjugate with respect to the multi-object observation likelihood and is closed under Chapman-Kolmogorov equation [15], [16]. In [15], the Labeled Multi-Bernoulli (LMB) RFS is employed to construct a multi-object filter which is able to produce track-valued estimates. We use LMB filter to estimate the states of the unknown number of targets. Also, we employed the parameters of LMB filter as the variables in the cost function introduced in [11] for the purpose of sensor resource allocation in sensor-control problem. Our simulation results confirm that our proposed method is more accurate than the state-of-art RFS-based sensor-control methods even for scenarios with high clutter rate.

The rest of the paper is organized as follows. In Sec. II an overview of the sensor-control framework is given. Then in Sec. III we briefly review the Labeled Multi-Bernoulli filter which is used to address the underlying multi-target state estimation problem. Section IV is dedicated to describe the defined cost function and implementation of Labeled Multi-Bernoulli sensor-control. Numerical results are presented in section V. Section VI conclude the paper.

## II. PROBLEM STATEMENT: SENSOR-CONTROL

Following [9], we formulate the sensor-control problem in the POMDP framework. The POMDP is a generalized form of Markov decision process (MDP) [19] in which there is no direct access to the *states* and the states information are only realized by noisy observations. The elements of the POMDP formulation in this paper are: a finite set of single-object state denoted by $X_k$, a set of sensor-control commands denoted by $\mathbb{S}$, a stochastic model for single-target state transition, a finite set of observations denoted by $Z$, a stochastic measurement model, and a cost function $\mathcal{V}(s; X)$ that returns a reward or cost for transition of the multi-object state to $X$ via applying an action command $s \in \mathbb{S}$.

The purpose of sensor-control is to find the control command $\breve{s} \in \mathbb{S}$ which minimizes the defined cost function. In stochastic filtering, where the multi-target states $X_{k-1}$ and $X_k$ are characterized by their distributions, the control command $\breve{s}$ is commonly chosen to minimize the statistical mean of the cost function $\mathcal{V}(s; X)$ over all observations,

$$\breve{s} = \underset{s \in \mathbb{S}}{\operatorname{argmin}} \left\{ \mathbb{E}_{Z(s)} \left[ \mathcal{V}(s; X) \right] \right\}. \quad (1)$$

In POMDP, a Bayesian filtering scheme is commonly utilized as the framework to formulate target evolution. The latest development in multi-target Bayesian filtering is the Generalized Labeled Multi-Bernoulli (GLMB) filter [16] and its special case the Labeled Multi-Bernoulli (LMB) filter [15]. The LMB filter is a solution to the multi-object Bayes filter [20] and it produces track-valued estimates in a mathematically principled manner [15]. In terms of accuracy of estimation, the LMB filter outperforms the multi-Bernoulli filter. The main reason is that the LMB filter uses less approximations than the multi-Bernoulli filter [15]. Indeed, the LMB only involves one approximation of the posterior density, while the multi-Bernoulli filter requires two approximations on the multi-target posterior probability generating functional [15]. Also, the performance of LMB filter in terms of OSPA error values is similar to the $\delta$-GLMB filter [15] which has already proved to outperform the PHD and CPHD filters [15], [16], [21]. The reason lies in the fact that the $\delta$-GLMB filter propagates a parameter approximation of the multi-object posterior, whereas the PHD and CPHD filters are the first moment approximation to the multi-target Bayes filter.

Due to the above mentioned advantages of LMB filter, this paper focuses on an effective sensor-control solution for LMB filter using measurements of controlled sensors. The task-driven approach to sensor-control within multi-target filtering schemes is to choose the cost function in terms of the predicted multi-target state and the expected update outcomes for every admissible control command. Before we present our choice of cost function, the LMB filter is briefly reviewed in the next section – see [16] for details.

## III. LABELED MULTI-BERNOULLI FILTER

In this section the summery of the notion and formulation of the Labeled Multi-Bernoulli filter, which was introduced in [15], is presented. The notion of Labeled Multi-Bernoulli (LMB) RFS was introduced for the first time in [16]. LMB is a special case of generalized labeled multi-Bernoulli RFS. It is shown that LMB RFS is a conjugate prior with respect to the multi-object observation likelihood, and it is closed under the multi-target Chapman-Kolmogorov equation [15], [16].

In the following, we adopt the same notation used in [15] where the single-object states are denoted by lower-case letters, e.g. $x$, $\mathbf{x}$ and multi-object states by upper-case letters, e.g. $X$, $\mathbf{X}$. In order to distinguish between labeled and unlabeled states and their distributions, the labeled one is shown by bolded letters e.g. $\mathbf{x}$, $\mathbf{X}$, etc, spaces by blackboard bold e.g. $\mathbb{X}$, $\mathbb{L}$, $\mathbb{C}$, etc, and the class of finite subsets of a space $\mathbb{X}$ by $\mathcal{F}(\mathbb{X})$. Following [15], throughout the paper, the standard inner product notation is used and denoted by

$$\langle f, g \rangle \triangleq \int f(x)g(x)dx,$$

the generalized Kronecker delta is denoted by

$$\delta_Y(X) \triangleq \left\{ \begin{array}{ll} 1, & \text{if } X = Y \\ 0, & \text{otherwise} \end{array} \right.,$$

and the inclusion function, a generalization of the indicator function, by

$$1_Y(X) \triangleq \left\{ \begin{array}{ll} 1, & \text{if } X \subseteq Y \\ 0, & \text{otherwise} \end{array} \right..$$

The multi-object distribution of a GLMB RFS with state $\mathbf{X}$ and discrete label space $\mathbb{L}$ is given by

$$\pi(\mathbf{X}) = \Delta(\mathbf{X}) \sum_{c \in \mathbb{C}} w^{(c)}(\mathcal{L}(\mathbf{X})) \left[ p^{(c)} \right]^{\mathbf{X}}, \quad (2)$$

where

$$\Delta(\mathbf{X}) = \delta_{|\mathbf{X}|}(|\mathcal{L}(\mathbf{X})|)$$

and $\mathbb{C}$ is a discrete index set and $w^{(c)}(L)$ is the non-negative weights that only depends on the labels of multi-object state and satisfies $\sum_{L \subseteq \mathbb{L}} \sum_{c \in \mathbb{C}} w^c(L) = 1$. Each $p^{(c)}(x, \ell)$ is a probability density and satisfies $\int p^{(c)}(x, \ell)dx = 1$. In (2), $h^X \triangleq \prod_{x \in X} h(x)$, denotes the multi-object exponential, where $h$ is a real-valued function, with $h^\varnothing = 1$ by convention. Thus, the multi-object distribution of a GLMB RFS presented in (2), can be interpreted as a mixture of multi-object exponentials. Each term in this mixture consists of a weight $w^{(c)}$ that only depends on the labels of the multi-object state, and a multi-object exponential $[p^{(c)}]^{\mathbf{X}}$ that depends on the entire multi-object state. The projection $\mathcal{L}: \mathbb{X} \times \mathbb{L} \to \mathbb{L}$ is given by $\mathcal{L}(x, \ell) = \ell$ and $\mathcal{L}(\mathbf{X}) = \{\mathcal{L}(\mathbf{x}) : \mathbf{x} \in \mathbf{X}\}$ is the set of object labels of $\mathbb{X}$. A labeled RFS with state space $\mathbb{X}$ and discrete label space $\mathbb{L}$ is an RFS on $\mathbb{X}\times\mathbb{L}$ such that each realization has distinct labels [16], [21].

The LMB RFS is a special case of GLMB RFS and similar to the multi-Bernoulli RFS it is completely described by its components $\pi = \{(r^{(\zeta)}, p^{(\zeta)}) : \zeta \in \Psi\}$. The LMB RFS density is given by

$$\pi(\mathbf{X}) = \Delta(\mathbf{X})w(\mathcal{L}(\mathbf{X}))[p]^{\mathbf{X}}, \quad (3)$$

where

$$p(x,\ell) = p^{(\ell)}(x) \quad (4)$$

$$w(L) = \prod_{i\in\mathbb{L}}\left(1-r^{(i)}\right)\prod_{\ell\in L}\frac{1_{\mathbb{L}}(\ell)r^{(\ell)}}{(1-r^{(\ell)})} \quad (5)$$

comprising a single component [15].

Similar to the general multi-target Bayes filter, the LMB multi-target Bayes recursion propagates multi-target posterior density at each time according to the Chapman-Kolmogorov (prediction step) and the Bayes rule (update step).

*A. Prediction*

Reuter et al. [15] proved that a LMB RFS is closed under the Chapman-Kolmogorov equation which means if the current multi-object posterior is of the form of LMB, then the predicted multi-object distribution is still LMB. Assume that the prior and birth labeled multi-Bernoulli sets are modelled as follows:

$$\pi(\mathbf{X}) = \Delta(\mathbf{X})w(\mathcal{L}(\mathbf{X}))[p]^{\mathbf{X}} \quad (6)$$
$$\pi_B(\mathbf{X}) = \Delta(\mathbf{X})w_B(\mathcal{L}(\mathbf{X}))[p_B]^{\mathbf{X}} \quad (7)$$

where

$$w(L) = \prod_{i\in\mathbb{L}}\left(1-r^{(i)}\right)\prod_{\ell\in L}\frac{1_{\mathbb{L}}(\ell)r^{(\ell)}}{1-r^{(\ell)}}, \quad (8)$$

$$w_B(L) = \prod_{i\in\mathbb{B}}\left(1-r_B^{(i)}\right)\prod_{\ell\in L}\frac{1_{\mathbb{B}}(\ell)r_B^{(\ell)}}{1-r_B^{(\ell)}}, \quad (9)$$

$$p(x,\ell) = p^{(\ell)}(x) \quad (10)$$
$$p_B(x,\ell) = p_B^{(\ell)}(x). \quad (11)$$

with state space $\mathbb{X}$ and label space $\mathbb{L}_+ = \mathbb{B}\cup\mathbb{L}$ and with the condition $\mathbb{B}\cap\mathbb{L}=\varnothing$. The predicted multi-object distribution is then a LMB RFS and given by

$$\pi_+(\mathbf{X}) = \Delta(\mathbf{X})w_+(\mathcal{L}(\mathbf{X}))[p_+]^{\mathbf{X}} \quad (12)$$

where

$$w_+(I_+) = w_S(I_+\cap\mathbb{L})w_B(I_+\cap\mathbb{B}) \quad (13)$$

$$w_S(L) = \left(1-r^{(\cdot)}\eta_S(\cdot)\right)^{\mathbb{L}}\left(\frac{r^{(\cdot)}\eta_S(\cdot)}{1-r^{(\cdot)}\eta_S(\cdot)}\right)^L, \quad (14)$$

$$\eta_S(\ell) = \langle p_S(\cdot,\ell), p(\cdot,\ell)\rangle \quad (15)$$
$$p_+(x,\ell) = 1_{\mathbb{L}}(\ell)p_{+,S}(x,\ell) + 1_{\mathbb{B}}(\ell)p_B(x,\ell) \quad (16)$$
$$p_{+,S}(x,\ell) = \frac{\langle p_S(\cdot,\ell)f(x|\cdot,\ell), p(\cdot,\ell)\rangle}{\eta_S(\ell)} \quad (17)$$

where $p_S(\cdot|\ell)$ is the survival probability of an object and $f(x|\cdot,\ell)$ is the single-object transition model. Thus, if the multi-target posterior density is an LMB RFS with parameter set $\pi = \{(r^{(\ell)}, p^{(\ell)}) : \ell \in \mathbb{L}\}$ with state space $\mathbb{X}$ and label space $\mathbb{L}$ and the birth model is also an LMB RFS with parameter set $\pi_\mathbf{B} = \{(r_B^{(\ell)}, p_B^{(\ell)}) : \ell \in \mathbb{B}\}$ with state space $\mathbb{X}$ and label space $\mathbb{B}$ then the predicted multi-target density is also an LMB RFS with state space $\mathbb{X}$ and label space $\mathbb{L}_+ = \mathbb{B}\cup\mathbb{L}(\mathbb{B}\cap\mathbb{L}=\varnothing)$ and it is given by

$$\pi_+ = \{(r_{+,S}^{(\ell)}, p_{+,S}^{(\ell)}) : \ell\in\mathbb{L}\} \cup \{(r_B^{(\ell)}, p_B^{(\ell)}) : \ell\in\mathbb{B}\} \quad (18)$$

where

$$r_{+,S}^{(\ell)} = \eta_S(\ell)r^{(\ell)}, \quad (19)$$

$$p_{+,S}^{(\ell)} = \frac{\langle p_S(\cdot,\ell)f(x|\cdot,\ell), p(\cdot,\ell)\rangle}{\eta_S(\ell)}, \quad (20)$$

for more details see – [15] – proposition 2.

*B. Update*

In update step, if the multi-object density is an LMB RFS, then the multi-object posterior is not necessarily still an LMB RFS. Similar to multi-Bernoulli RFS [14], Reuter et al. [15] approximate the updated LMB RFS by its first moment. Thus, if the predicted multi-target density is an LMB RFS with parameter set $\pi_+ = \{(r_+^{(\ell)}, p_+^{(\ell)}) : \ell \in \mathbb{L}_+\}$, the multi-target posterior is then given by

$$\pi(\cdot|Z) = \{(r^{(\ell)}, p^{(\ell)}(\cdot) : \ell\in\mathbb{L}_+\} \quad (21)$$

where

$$r^{(\ell)} = \sum_{(I_+,\theta)\in\mathcal{F}(\mathbb{L}_+)\times\Theta_{I_+}} w^{(I_+,\theta)}(Z)1_{I_+}(\ell), \quad (22)$$

$$p^{(\ell)}(x) = \frac{1}{r^{(\ell)}}\sum_{(I_+,\theta)\in\mathcal{F}(\mathbb{L}_+)\times\Theta_{I_+}} w^{(I_+,\theta)}(Z)1_{I_+}(\ell)p^{(\theta)}(x,\ell), \quad (23)$$

where $\Theta_{I_+}$ denotes the space of mapping $\theta : I_+ \to \{0,1,\ldots,|Z|\}$ and,

$$w^{(I_+,\theta)}(Z) \propto w_+(I_+)\left[\eta_Z^{(\theta)}\right]^{I_+} \quad (24)$$

$$p^{(\theta)}(x,\ell|Z) = \frac{p_+(x,\ell)\psi_Z(x,\ell;\theta)}{\eta_Z^{(\theta)}(\ell)}, \quad (25)$$

$$\eta_Z^{(\theta)}(\ell) = \langle p_+(\cdot,\ell), \psi_Z(\cdot,\ell;\theta)\rangle, \quad (26)$$

$$\psi_Z(x,\ell;\theta) = \delta_0(\theta(\ell))q_D(x,\ell)$$
$$+ (1-\delta_0(\theta(\ell)))\frac{p_D(x,\ell)g(z_{\theta(\ell)}|x,\ell)}{\kappa(z_{\theta(\ell)})} \quad (27)$$

where, $g(z|x)$ is the single-sensor measurement likelihood, $p_D(\cdot,\ell)$ denotes probability of detection, $q_D(\cdot,\ell) = 1-p_D(\cdot,\ell)$ is the probability of a missed detection, and $\kappa(\cdot)$ intensity function of the Poisson distributed clutter process.

*C. Implementation*

Details of sequential Monte Carlo implementation of the LMB filter are presented in [15]. In the implementation, the number of hypotheses grows exponentially. For computational reduction, targets and measurements are subjected to spatial grouping and gating and the update step is run in parallel for those groups. In order to keep only the most significant hypotheses, several methods of truncation are proposed in the literature [15], [16], [21].

In the prediction step, the $K$-shortest path algorithm is used to truncate the predicted LMB without computing all the prediction hypotheses and their weights [22]. To avoid computing all the hypotheses and their weights in the update step, the updated LMB multi-target posterior is truncated, via the ranked assignment algorithm. Murty's method is employed for the ranked assignment process in which only the $M$ most significant association hypotheses are evaluated [23]. For more details see [15].

## IV. LABELED MULTI-BERNOULLI SENSOR-CONTROL

As it was mentioned earlier, the POMDP approach to the sensor-control problem comprises of a multi-target tracking framework and a stochastic decision making solution to choose the optimal command via an objective function. Due to estimation accuracy of the LMB filter, in this study we choose LMB filter to carry out the multi-target tracking problem. The only drawback of this filter compare to the other RFS-based filters is the computational complexity of its update step.

To reduce the complexity of our proposed method, in the sensor-control step, instead of using the update formulation of the LMB filter, the multi-Bernoulli update [14] is employed. In order to use the update step of the multi-Bernoulli filter, first the predicted parameters of the LMB are computed and the augmented label state is discarded. Note that the unlabeled version of the LMB parameters is equal to the multi-Bernoulli parameters. Having the predicted parameters of the multi-Bernoulli distribution, the number and states of the targets are pre-estimated. For each sensor-control command, a set of pseudo-measurements are generated (according to the pre-estimated targets) using the Predicted Ideal Measurement Set (PIMS) approach [17], then the multi-Bernoulli update is performed. By acquiring posterior multi-Bernoulli densities for each admissible command, the command that maximizes the utility of the measurement is chosen and applied. After changing the state of the sensor(s) and receiving the actual set of measurements, the LMB update is performed. The main steps of the sensor-control with LMB filter are given in Algorithm 1.

### A. Cost function

The most common approach to choose the optimal control command in the sensor-control solutions are based on maximizing an information theoretic reward function such as Rényi divergence [7], [8], [24]. The main rationale behind choosing such reward functions is that the information encapsulated by the estimated multi-target distribution is expected to gradually increase as further measurements become available by time.

Following our preliminary study [9], we take a different approach in which the updated parameters of the multi-Bernoulli filter are used to define a new cost function. Note that the multi-Bernoulli parameters are updated by extracting the unlabeled version of the predicted LMB parameters. In the sensor-control step, these parameters are then updated by using the update formulation of the multi-Bernoulli filter. Our approach is to consider a cost function that quantifies the average uncertainty in all possible multi-target state estimates after each update step. This cost is not totally independent of the prediction outcomes, and state estimates extracted from

**Algorithm 1** The LMB multi-target filtering recursion with sensor-control.

INPUTS: dynamic model $f(x|\cdot,\ell)$, LMB birth model parameters, prior LMB parameters from time $k-1$, detection probability $p_D(\cdot)$, measurement likelihood function $g_k(\cdot|x)$, and clutter intensity $\nu(\cdot)$ and its integral $\lambda_c$, current sensor(s) location(s), finite set of admissible sensor-control commands $\mathbb{S}$.
OUTPUT: The best control command $\overset{\star}{s}$ and updated LMB parameters.

**Prediction:**
1: Compute the predicted LMB component parameters.
2: Extract unlabeled version of LMB parameters.

**Pre-estimation:**
3: Compute the prediction estimates of the number and states of objects.

**Sensor-control:**
4: **for** $s \in \mathbb{S}$ **do**
5:     Construct the PIMS, $\mathring{Z}(s)$.
6:     Update the multi-Bernoulli distribution parameters.
7:     Compute the cost $\mathcal{V}(s; X)$
8: **end for**
9: $\overset{\star}{s} \leftarrow \arg\min_s \mathcal{V}(s; X)$

**Measurement:**
10: Apply the control command $\overset{\star}{s}$ to change the sensor state
11: Collect the actual measurements from controlled sensor(s).

**Update:**
12: Use the measurement set to update the LMB parameters.

predicted multi-Bernoulli density are used to calculate the proposed cost function. The main difference here is that our focus is on the quality of the updated density in terms of level of uncertainties, not the information gained from prediction to update (e.g. Rényi divergence function).

The updated distribution depends on the receiving measurement. The generated measurements set is also a function of the chosen sensor control command. In principle the whole distribution of all possible measurement sets is used to compute the update distribution. However, to reduce the computational complexity, we adopt the predicted ideal measurement set (PIMS) [17] for the purpose of updating the multi-target distribution and computing the cost from it. To define the new cost function, we note that the predicted ideal measurement set depends on the chosen control command. For each command, we first compute the PIMS, then calculate an updated multi-object distribution by considering the PIMS as the acquired measurement. A linear combination of the normalized errors of the number of targets and their estimated states is considered as a measure of uncertainty associated with estimation of the multi-target state and as the cost function:

$$\mathcal{V}(s; X) = \eta \; \varepsilon^2_{|X|}(s) + (1-\eta) \; \varepsilon^2_X(s), \quad (28)$$

where $\varepsilon^2_{|X|}(s)$ denotes the normalized error of estimated cardinality of the multi-target state, $\varepsilon^2_X(s)$ denotes the normalized error of the multi-target state estimate, and $\eta \in [0,1]$ is a user-defined constant parameter to tune the influence of the error terms on the total sensor control cost. Appearance of the $X_{k-1}$ as an argument of the cost function is to emphasize that the cost not only depends on the selected control command, but also on the prior distribution. It is important to note that the expectation term in (1) does not appear as we use the *predicted ideal measurement set* (PIMS) approach [17] instead of sampling and averaging in measurement space. The details of computing the PIMS, and defining and computing the normalized error terms, $\varepsilon^2_{|X|}(s)$ and $\varepsilon^2_X(s)$ for SMC implementation are presented in Sec. IV-B.

The quality of sensor measurements usually depends on a sensor state (e.g. the sensor location) which is assumed to be controllable, and the sensor-control problem is focused on choosing the command that would lead to the best sensor state.

*B. Implementation*

Suppose that at the time $k-1$, the multi-target distribution is modelled by a LMB RFS with parameters $\pi = \{(r^{(\ell)}, p^{(\ell)}) : \ell \in \mathbb{L}\}$ and $\pi_\mathbf{B} = \{(r_B^{(\ell)}, p_B^{(\ell)}) : \ell \in \mathbb{B}\}$, in which each single target density $p^{(\ell)}(\cdot)$ is represented by a set of weighted samples $\{(\omega_i^{(\ell)}, x_i^{(\ell)})\}_{i=1}^{J^{(\ell)}}$ and the birth density $p_B^{(\ell)}(\cdot)$ is represented by $\{(\omega_{B,i}^{(\ell)}, x_{B,i}^{(\ell)})\}_{i=1}^{B^{(\ell)}}$. In the prediction step, the LMB filter propagates the LMB components based on the temporal information from the transition density, the probability of survival, and the predefined LMB birth terms. The predicted LMB density is denoted by $\pi_+ = \{(r_+^{(\ell)}, p_+^{(\ell)}) : \ell \in \mathbb{L}_+\}$ where each single target density $p_+^{(\ell)}(\cdot)$ is represented by a set of weighted samples $\{(\omega_{+,i}^{(\ell)}, x_{+,i}^{(\ell)})\}_{i=1}^{J_+^{(\ell)}}$.

For each label $\ell \in \mathbb{L}_+$, if the probability of existence $r_+^{(\ell)}$ is greater than a user-defined threshold (chosen at 0.5 in our simulation studies), the EAP estimate of a single-object state is computed as follows:

$$\hat{x}^{(\ell)} = \sum_{i=1}^{J_+^{(\ell)}} \omega_{+,i}^{(\ell)} \, x_{+,i}^{(\ell)}. \quad (29)$$

Each of the above estimates represent a predicted target. Following the PIMS approach [17], an ideal set of measurements are then generated from the predicted target state estimates. This hypothetical set of measurement would depend on not only the predicted number and states of targets, but also the new state of the sensor(s) after a sensor command is applied. Indeed, for each possible sensor control command $s \in \mathbb{S}$, a different set of ideal measurements, $\mathring{Z}(s)$, is computed. Considering this set as the actual measurement set, we can now run the update step and calculate the cost corresponding to that command.

We note that the LMB prediction step "actually coincides with performing the prediction on the unlabeled process and interpreting the component indices as track labels."–Remark 3 from [15]. Therefore, we can simply remove the labels from the predicted LMB multi-object state, and update the existence probabilities and density particles and weights through the CB-MeMBer update step [14] in which $\mathring{Z}(s)$ is taken as the actual measurement set. It is important to note that this update step needs to be repeated for each hypothetical measurement set $\mathring{Z}(s)$. By using the CB-MeMBer update on the unlabeled components we avoid to repeatedly run the computationally expensive update of LMB filter. Hence, substantial savings are achieved in terms of computational cost of our sensor-control method.

**Computing the cost:** The cost defined in (28) comprises two normalized error terms, $\varepsilon_{|X|}^2(s)$ as the error for the cardinality estimate, and $\varepsilon_X^2(s)$ as the error for the multi-target state estimate. Both terms depend on the updated multi-object posterior which in turn depends on the PIMS computed for the command $s$. Assume that for each control-command $s \in \mathbb{S}$, the CB-MeMBer updated *unlabeled* multi-Bernoulli is given by $\{r^{(i)}(s), p^{(i)}(s, \cdot)\}_{i=1}^{M(s)}$ where each single Bernoulli density $p^{(i)}(s, \cdot)$ is approximated by particles $\{\omega_j^{(i)}(s), x_j^{(i)}(s)\}_{j=1}^{J^{(i)}(s)}$.

We choose and calculate the statistical expectation of the cardinality variance as a meaningful measure for its estimation error. In terms of the updated probabilities of existence, it is given by:

$$\sigma_{|X|}^2(s) = \sum_{i=1}^{M(s)} \left[ r^{(i)}(s)(1 - r^{(i)}(s)) \right]. \quad (30)$$

The above given value is maximum when $\forall i, r^{(i)}(s) = 0.5$ which leads to $\max\{\sigma_{|X|}^2(s)\} = \frac{M(s)}{4}$. Thus, the normalized cardinality error term can be computed as follows:

$$\varepsilon_{|X|}^2(s) = \frac{4\sigma_{|X|}^2(s)}{M(s)}. \quad (31)$$

To arrive at a meaningful measure for the normalized state estimation error term $\varepsilon_X^2(s)$ in the cost defined in (28), we consider the following total state estimation error:

$$\varepsilon_X^2(s) = \sum_{i=1}^{M(s)} \left[ r^{(i)}(s) \epsilon_{x^{(i)}}^2(s) \right] \, / \, \sum_{i=1}^{M(s)} r^{(i)}(s) \quad (32)$$

which is the weighted average of normalized estimation errors of the states of single targets associated with each single Bernoulli component. Before we present how the normalized error terms $\epsilon_{x^{(i)}}^2(s)$ are computed, we note that the averaging weights are the updated probabilities of existence. The rationale behind this choice of weights is that Bernoulli components with larger probabilities of existence contribute more strongly to the EAP estimate of the multi-object state –see section IV-A.4 in [14].

To compute the normalized single Bernoulli component errors $\epsilon_{x^{(i)}}^2(s)$, we first formulate the actual error denoted by $\varsigma_{x^{(i)}}^2(s)$, then its maximum, upon which a normalized measure will be given by $\varsigma_{x^{(i)}}^2(s)/\max \varsigma_{x^{(i)}}^2(s)$. In practice, we are commonly interested in minimizing the estimation error of *selected elements* of target states. For instance, in some applications, the prime interest is in *location*, and target speed is included in the single-target state vector due to its appearance in motion and perhaps measurement models. In such target-tracking applications, an intuitive scalar measure for the single Bernoulli component error is given by the product of the variances of the target location coordinates. If the stochastic variations of target location coordinates are independent, this measure will translate into the absolute determinant of the covariance matrix of the target location.

In case of tracking multiple-targets in 2D space, the single Bernoulli component error term, $\varsigma_{x^{(i)}}^2(s)$, is given by:

$$\varsigma_{x^{(i)}}^2(s) = \sigma_{\mathrm{x}^{(i)}}^2(s) \, \sigma_{\mathrm{y}^{(i)}}^2(s) \quad (33)$$

where x and y denote the x and y-coordinates of the single-target location (part of its state vector $x$). Having the updated

particles and weights of each Bernoulli component, the single-coordinate errors can be calculated as follows:

$$\sigma^2_{x^{(i)}}(s) = \sum_{j=1}^{J^{(i)}(s)} \omega_j^{(i)}(s) \left(x_j^{(i)}(s)\right)^2 - \left(\sum_{j=1}^{J^{(i)}(s)} \omega_j^{(i)}(s) x_j^{(i)}(s)\right)^2$$
$$\sigma^2_{y^{(i)}}(s) = \sum_{j=1}^{J^{(i)}(s)} \omega_j^{(i)}(s) \left(y_j^{(i)}(s)\right)^2 - \left(\sum_{j=1}^{J^{(i)}(s)} \omega_j^{(i)}(s) y_j^{(i)}(s)\right)^2 \quad (34)$$

where $x_j^{(i)}(s)$ and $y_j^{(i)}(s)$ denote the coordinates extracted from the particle $x_j^{(i)}(s)$ and power operation is element-wise operation. To normalize the total state estimation error term $\varsigma^2_{x^{(i)}}(s)$ in (33), we note that with equally weighted particles, i.e. when $\forall j, \omega_j^{(i)}(s) = 1/J^{(i)}(s)$, the particles representing the $i$-th single Bernoulli component do not convey any information and the above estimation variances adopt their maximum values as follows:

$$\max\{\sigma^2_{x^{(i)}}(s)\} = \tfrac{1}{J^{(i)}(s)}(1 - \tfrac{1}{J^{(i)}(s)}) \sum_{j=1}^{J^{(i)}(s)} \left(x_j^{(i)}(s)\right)^2$$
$$\max\{\sigma^2_{y^{(i)}}(s)\} = \tfrac{1}{J^{(i)}(s)}(1 - \tfrac{1}{J^{(i)}(s)}) \sum_{j=1}^{J^{(i)}(s)} \left(y_j^{(i)}(s)\right)^2. \quad (35)$$

Thus, the single Bernoulli error terms $\varsigma^2_{x^{(i)}}(s)$ in (33) can be normalized as follows:

$$\epsilon^2_{x^{(i)}}(s) = \frac{\sigma^2_{x^{(i)}}(s)\ \sigma^2_{y^{(i)}}(s)}{\max\{\sigma^2_{x^{(i)}}(s)\}\ \max\{\sigma^2_{y^{(i)}}(s)\}} \quad (36)$$

and the computed values can be used in (32) to calculate the normalized state estimation error term in the cost.

Having the cost values computed for all admissible sensor control commands, the best command $\overset{\star}{s}$ is then chosen as the one incurring the smallest cost:

$$\overset{\star}{s} = \underset{s \in \mathbb{S}}{\operatorname{argmin}} \mathcal{V}(s; X). \quad (37)$$

As the cost function of the proposed method is a combination of posterior expected errors of cardinality and states, henceforward, we call it the *Posterior Expected Error of Cardinality and States* (PEECS).

## V. NUMERICAL STUDIES

To demonstrate the performance of our method with measurements that guarantee full observability of the targets, we have run a case study involving a complex scenario. In this scenario, we choose a non-linear nearly-constant turn model reported in [14]. In this case, each single target state $x = [\bar{x}^\top\ \omega]^\top$ is comprised of location and velocity in Cartesian coordinates, denoted by $\bar{x} = [x\ y\ \dot{x}\ \dot{y}]^\top$ and turning rate, denoted by $\omega$. The state dynamics are given by:

$$\bar{x}_k = F(\omega_{k-1})\bar{x}_{k-1} + G\epsilon_{k-1},$$
$$\omega_k = \omega_{k-1} + T\gamma_{k-1},$$

where

$$F(\omega) = \begin{bmatrix} 1 & 0 & \tfrac{\sin \omega T}{\omega} & -\tfrac{1-\cos \omega T}{\omega} \\ 0 & 1 & \tfrac{1-\cos \omega T}{\omega} & \tfrac{\sin \omega T}{\omega} \\ 0 & 0 & \cos \omega T & -\sin \omega T \\ 0 & 0 & \sin \omega T & \cos \omega T \end{bmatrix}, G = \begin{bmatrix} \tfrac{T^2}{2} & 0 \\ 0 & \tfrac{T^2}{2} \\ T & 0 \\ 0 & T \end{bmatrix},$$

$T = 1\,\text{s}$, $\epsilon_{k-1} \sim \mathcal{N}(\cdot; 0, \sigma^2_\epsilon I)$, $\sigma_\epsilon = 15\,\text{m/s}^2$, and $\gamma_{k-1} \sim \mathcal{N}(\cdot; 0, \sigma^2_\gamma I)$, $\sigma_\gamma = (\pi/180)\,\text{rad/s}$. The birth RFS is a multi-Bernoulli with density $\pi_\Gamma = \{(r_\Gamma^{(i)}, p_\Gamma^{(i)})\}_{i=1}^4$ where $r_\Gamma^{(1)} = r_\Gamma^{(2)} = 0.02$, $r_\Gamma^{(3)} = r_\Gamma^{(4)} = 0.03$ and $p_\Gamma^{(i)}(x) = \mathcal{N}(x; m_\gamma^{(i)}, P_\gamma)$ where

$$\begin{aligned} m_\gamma^{(1)} &= [\text{-}1500\ 0\ 250\ 0\ 0]^\top, \\ m_\gamma^{(2)} &= [\text{-}250\ 0\ 1000\ 0\ 0]^\top, \\ m_\gamma^{(3)} &= [\ 250\ 0\ 750\ 0\ 0]^\top, \\ m_\gamma^{(4)} &= [\ 1000\ 0\ 1500\ 0\ 0]^\top, \\ P_\gamma &= \operatorname{diag}(50^2, 50^2, 50^2, 50^2, (6 \times \tfrac{\pi}{180})^2). \end{aligned}$$

The sensor can detect an object in location $\mathbf{o} = [x_o\ y_o]^\top$ with the following probability that depends on the location of both the sensor and object locations:

$$p_D(\mathbf{s}, \mathbf{o}) = \begin{cases} 1, & \text{if } \|\mathbf{o} - \mathbf{s}\| \leq R_0 \\ \max\{0, 1 - \mathfrak{h}(\|\mathbf{o} - \mathbf{s}\| - R_0)\}. & \text{otherwise} \end{cases} \quad (38)$$

Overall, there are five objects in the surveillance area, positioned relatively close to each other. Their initial state vectors are: $[800\ 600\ 1\ 0]^\top$, $[650\ 500\ 0.3\ 0.6]^\top$, $[620\ 700\ 0.25\ 0.45]^\top$, $[750\ 800\ 0\ 0.6]^\top$, and $[700\ 700\ 0.2\ 0.6]^\top$, where the units of x and y are meters and $\dot{x}$ and $\dot{y}$ are m/s. The objects move according to the constant velocity model.

Each measurement includes a set of ranges and bearings, and the observation model is given by:

$$z_k = \left[\arctan(\tfrac{x_k}{y_k})\quad \sqrt{x_k^2 + y_k^2}\right]^\top + \zeta_k,$$

where $\zeta_k \sim \mathcal{N}(\cdot; 0, R_k)$ is the measurement noise with covariance $R_k = \operatorname{diag}(\sigma^2_\theta, \sigma^2_r)$ in which the scales of range and bearing noise are $\sigma_\theta = (\pi/180)\,\text{rad}$ and $\sigma_r = 5\,\text{m}$. The clutter RFS follows the uniform Poisson model over the surveillance region $[\text{-}\pi/2\,,\pi/2]\,\text{rad} \times [0, 2000]\,\text{m}$, with $\lambda_c = 1.6 \times 10^{-3}\,(\text{rad m})^{-1}$.

The sensor initial position is at x = 0 and y = 1500. Targets enter the scene with the following position, velocity, turning velocity, birth and death time:

$$\begin{aligned} x_{T_1} &= [1000\ \text{-}20\ 1500\ \text{-}20\ \tfrac{\pi}{180}]^\top, & k^b_{T_1} &= 1\ \ \ , & k^d_{T_1} &= 35, \\ x_{T_2} &= [\text{-}250\ 20\ 1000\ 3\ \tfrac{\pi}{270}]^\top, & k^b_{T_2} &= 5\ \ \ , & k^d_{T_2} &= 50, \\ x_{T_3} &= [\text{-}500\ 11\ 250\ 10\ \tfrac{\pi}{180}]^\top, & k^b_{T_3} &= 20\ , & k^d_{T_3} &= 50, \\ x_{T_4} &= [\text{-}500\ 14\ 250\ 0\ 0]^\top, & k^b_{T_4} &= 15\ , & k^d_{T_4} &= 50, \\ x_{T_5} &= [\ 250\ 11\ 750\ 5\ \tfrac{\pi}{360}]^\top, & k^b_{T_5} &= 10\ , & k^d_{T_5} &= 50. \end{aligned}$$

Intuitively, we expect the sensor start moving toward the targets and for each time step remains in vicinity of them. As it is shown in Fig. 1, sensor start moving from the initial position and as it was expected, after a few steps it remains among the manoeuvring targets. The estimation error are computed based on the Optimal SubPattern Assignment (OSPA) metrics introduced in [25] (cutoff parameter $c = 100$ and order parameter $p = 2$). The comparative averaged error performance of LMB PEECS sensor control and CB-MeMBer PEECS sensor control are shown in Fig. 2 over 200 Monte Carlo runs. Figure 2(a) and 2(b) show the cardinality and localization errors for both methods respectively. At first the cardinality and localization errors are high due to uncertainty in the number and states of targets. After the time step $k = 20$ both the cardinality and localization errors are fixed through the rest of the simulation. The LMB PEECS errors are comparatively lower than the CB-MeMBer PEECS error. The total OSPA error are shown in Fig. 2(c). The superiority of the LMB PEECS method is due to accuracy of the LMB filter which is the result of proper approximation in update step of the LMB filter. As it was mentioned in Sec. II, unlike multi-Bernoulli filter the LMB filter uses a more accurate update approximation. More precisely, the LMB filter uses a more acurate update approximation than the CB-MeMBer filter by exploiting the conjugate prior labeled RFSs [15].

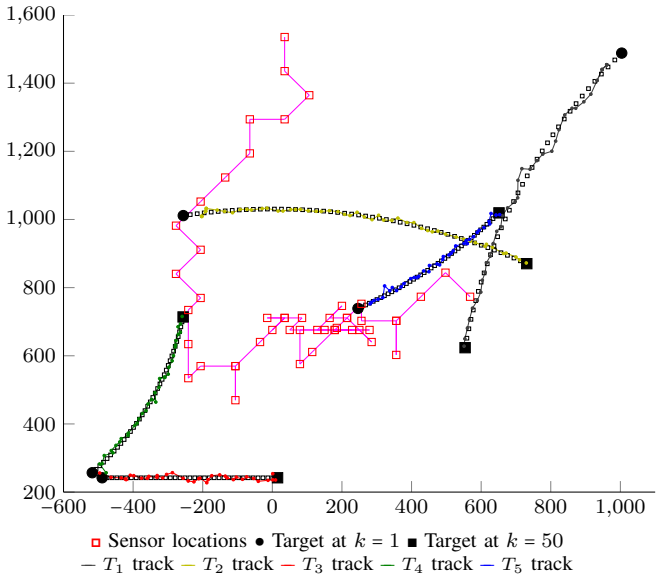

Fig. 1: Sensor and target locations during $k = 1, \ldots, 50$.

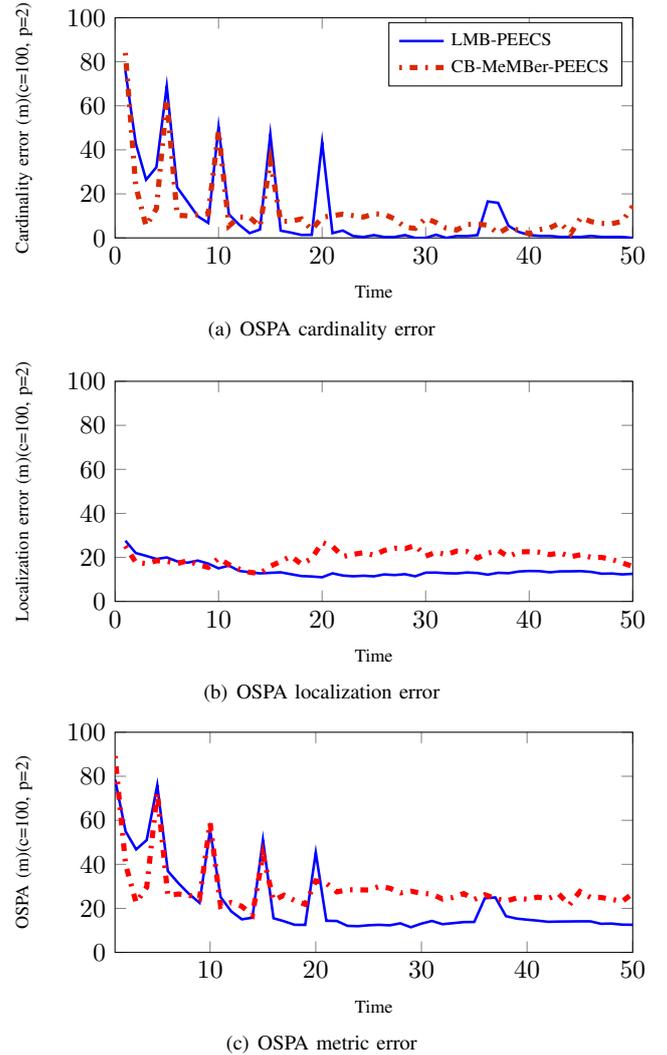

(a) OSPA cardinality error

(b) OSPA localization error

(c) OSPA metric error

Fig. 2: Estimation errors of the PEECS sensor control method.

## VI. CONCLUSIONS AND FUTURE STUDIES

A novel sensor-control method was proposed in this paper, for controlling mobile sensor(s) in such a way that minimum expected errors are achieved in a multi-target tracking application. The proposed method works based on choosing the sensor-control command that is expected to lead to the lowest cost, and the cost is defined and formulated in terms of cardinality and single-target state estimation errors. Implementation of the cost computation and its minimization within a labeled multi-Bernoulli filter was elaborated and a step-by-step algorithm was presented. In a challenging simulation scenario, our proposed method demonstrated success in optimal guidance of a mobile sensor in tracking of up to 5 targets which can appear and disappear in/from the scene. Compared to the scenario where a similar cost function was used within a CB-MeMBer filter, we showed that our new method developed for the labeled multi-Bernoulli filter performs better in terms of OSPA errors in the same challenging scenario. This can also be due to the advantageous nature of the labeled multi-Bernoulli filter in terms of its better accuracy in approximating the update step.

This work can be extended to applications where detection and tracking of *targets of interest* are required. In this case, a new task-driven cost function would be needed in which not only the estimation errors are considered but also the track labels produced by the labeled multi-Bernoulli filter are utilized. Such a sensor-control routine would guide the sensor(s) towards sensor states which are likely to lead to better estimates of the states of the targets of interest (with particular labels) only.

ACKNOWLEDGMENT

This work was supported by the ARC Discovery Project Grant DP130104404.